\documentclass[notitlepage]{revtex4-1} 
\usepackage{graphicx}

\usepackage{hyperref} 					 
\usepackage{listings} 					 
\usepackage{color}    					 
\usepackage{natbib}   					 
\usepackage{subfig} 					 

\usepackage{booktabs}
\usepackage[ansinew]{inputenc}				     
\usepackage{psfrag}
\usepackage{pstricks}
\usepackage{amsfonts}
\usepackage{amsmath}
\usepackage{bm}
\usepackage{lipsum}
\usepackage{pgf}
\usepackage{tikz}
\usetikzlibrary{patterns}

\usepackage{algorithm}
\usepackage{algorithmic}

\usepackage{array}
\usepackage{booktabs}
\newcolumntype{C}[1]{>{\centering\arraybackslash}m{#1}}

\usepackage{filecontents}

\usepackage{amsmath}
\usepackage[misc,geometry]{ifsym} 
\begin{document}
\title{Designing Photonic Topological Insulators with Quantum-Spin-Hall Edge States using Topology Optimization}

\author{Rasmus E. Christiansen$^1$}
\email{Corresponding email: raelch@mek.dtu.dk}

\author{Fengwen Wang$^1$}

\author{Ole Sigmund$^1$}

\author{S{\o}ren Stobbe$^2$}

\affiliation{{$^1$}Department of Mechanical Engineering, Solid Mechanics, Technical University of Denmark, Nils Koppels Allé, B.\ 404, DK-2800 Kgs. Lyngby, Denmark} 
\affiliation{{$^2$}Department of Photonics Engineering, DTU Fotonik, Technical University of Denmark, Building 343, DK-2800 Kgs.\ Lyngby, Denmark}


\begin{abstract}
\noindent Designing photonic topological insulators is highly non-trivial because it requires inversion of band symmetries around the band gap, which was so far done using intuition combined with meticulous trial and error. Here we take a completely different approach: we consider the design of photonic topological insulators as an inverse design problem and use topology optimization to maximize the transmission through an edge mode with a sharp bend. Two design domains composed of two different, but initially identical, C$_\text{6v}$-symmetric unit cells define the geometrical design problem. Remarkably, the optimization results in a photonic topological insulator reminiscent of the shrink-and-grow approach to quantum-spin-Hall photonic topological insulators but with notable differences in the topology of the crystal as well as qualitatively different band structures and with significantly improved performance as gauged by the band-gap sizes, which are at least 50 \% larger than previous designs. Furthermore, we find a directional beta factor exceeding 99 \%, and very low losses for sharp bends. Our approach allows for the introduction of fabrication limitations by design and opens an avenue towards designing PTIs with hitherto unexplored symmetry constraints.
\end{abstract}

\maketitle

\section{1. Introduction}
Defects in photonic crystals (PCs) such as cavities or waveguides allow for confining light to subwavelength dimensions \cite{Joannopoulos:08:Book} and has enabled a wealth of studies of light-matter interaction at the nanoscale \cite{PAP_Lodahl2015}. However, photonic crystals are also very sensitive to structural disorder, which results in substantial variations in the resonance frequencies and quality of PC cavities \cite{Minkov:13} and backscattering-induced Anderson localization of PC waveguides \cite{PAP_Sapienza2010}, which are serious impediments for applications of PCs. Photonic topological insulators (PTIs) provide a fundamentally different approach to confinement where light is confined as symmetry-protected edge states between PCs with different band topologies and such edge modes may be robust against backscattering \cite{LU_2014, KHANIKAEV_2017}.

While PTIs have been realized in several photonic platforms featuring broken time-reversal symmetry \cite{WANG_2009,KHANIKAEV_ET_AL_2012, RECHTSMAN_2013}, which allows unidirectional modes similar to the quantum Hall effect \cite{HASAN_KANE_2010}, these approaches require materials and geometries incompatible with inorganic semiconductors and planar technology, e.g., III-V materials for active communication devices or solid-state quantum optics and silicon photonics for chip-scale routing and optical interconnects. More recently, time-reversal-symmetric all-dielectric PC-based PTI designs for the photonic quantum-spin-Hall effect have been proposed \cite{Wu2015}, which was later extended to planar PTIs \cite{BARIK_2016} in which a number of chiral quantum optical effects \cite{PAP_Lodahl2017} have been demonstrated experimentally \cite{Barik_2018,Parappurath2018}. As an alternative approach, breaking parity symmetry leads to valley-Hall PTIs \cite{Ma2016,Shalaev2019}.

A key ingredient in PTI design is band inversion, i.e., the symmetry of the photonic bands must be swapped around the photonic band gap in order to induce the topological transition forming the edge state. This is in general a non-trivial task and thus remarkably few designs are available in the literature. The starting point for all planar time-reversal-symmetry-protected quantum-spin-Hall PTI designs so far is photonic graphene, which features a doubly degenerate Dirac point that upon suitable lattice perturbations may be gapped in topologically non-trivial ways. Notably, all dielectric photonic-crystal-based quantum-spin-Hall PTI designs to date employ the shrink-and-grow strategy first used by Wu and Xu \cite{Wu2015}. Here we take an entirely different approach: we consider the design of a PTI as an inverse numerical optimization problem seeking the maximum transmission of edge states between two crystals with C$_\text{6v}$ symmetry by manipulating the crystal geometry. We solve this problem using density-based topology optimization (TO) \cite{BOOK_TOPOPT_BENDSOE}, an optimization-based design method utilizing adjoint sensitivity analysis \cite{TORTORELLI_1994} and gradient-based optimization algorithms \cite{SVANBERG_2002} to efficiently solve design problems with potentially billions of degrees of freedom. Density-based topology optimization has previously been successfully applied across numerous areas, including recent optimization and design of complex structures and materials in mechanics \cite{AAGE_2017,YANG_2018,WANG_2018b,DONG_ET_AL_2016}, optics and photonics \cite{WANG_2018,SELL_ET_AL_2017,JENSEN_SIGMUND_2011,MOLESKY_2018}, thermo-fluidics \cite{ALEXANDERSEN_ET_AL_2018}, and acoustics \cite{CHRISTIANSEN_SIGMUND_2016,PARK_ET_AL_2014}. Very recently, TO was used to devise new designs of acoustic topological insulators \cite{CHRISTIANSEN_2019} and here we apply this approach for the first time to the design of PTIs. Remarkably, our approach generates topologically non-trivial bands that resemble those obtained from the use of the shrink-and-grow strategy although these appears spontaneously through the optimization process and not as a imposed design constraint. While our approach thus qualitatively reproduces concepts and ideas already available in the literature, the resulting unit cells are entirely different with non-trivial geometric features, which turn out to significantly improve figures of merit: The relative band gaps of the two crystal phases are $19\%$ and $6\%$, we find a directional beta-factor $>99\%$, and the bending losses are extremely small.

\section{2. Model}

\begin{figure}[h] 
	\centering
	{
		\includegraphics[width=0.47\textwidth]{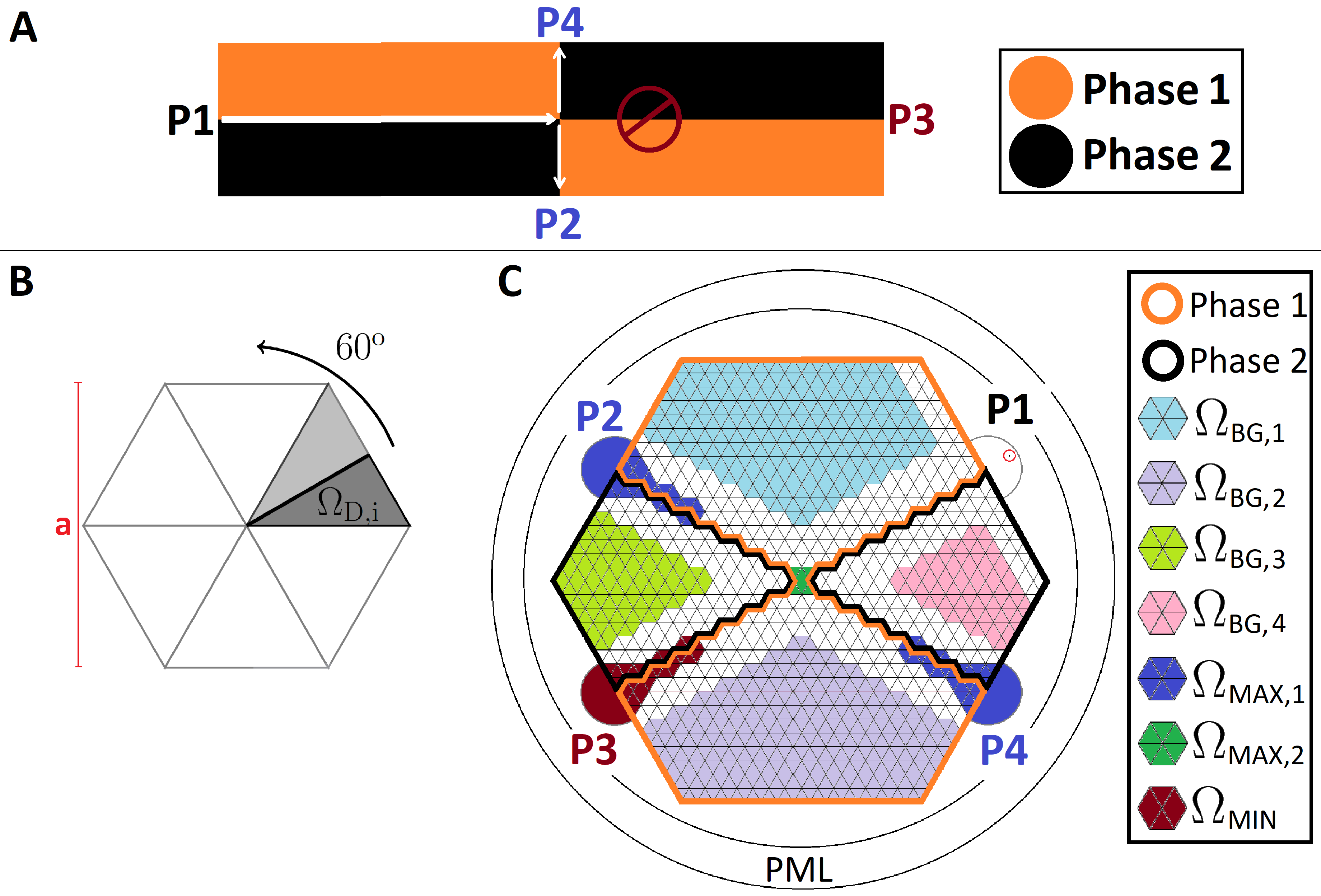}
		\caption{\textbf{Crystal Symmetry and Design Problem Configuration.} \textbf{(A)} Illustration of the expected transmission for a PTI in our design-domain topology. Light impinging from P1 reaches a crossing between the different crystal phases, and at the crossing, light cannot propagate forwards, it can only scatter to the left or right. \textbf{(B)} Symmetry constraints for a single unit cell, which in both phases is restricted to C$_\text{6v}$ symmetry. The designable part of the unit cell is denoted $\bm{\Omega}_{\text{D},i}$ (dark gray) and mirrored across the bold black line. The triangular region (light gray + dark gray) is replicated by 60${^{\text{o}}} \cdot n, n \in \lbrace 1,2,...,5 \rbrace$ rotations to fill the hexagonal unit cell. \textbf{(C)} The full optimization domain showing the PC phase distribution. Light is impinging on port P1 from an excitation source whose position is indicated with a black dot highlighted by a red circle at port P1 and the intensity at ports P2-P4 is recorded. The target function maximizes the intensity at P2 and P4 while minizing the intensity at P3. The regions containing the first and second crystal phases are outlined using orange and black denoted Phase 1 and Phase 2, respectively. A number of subdomains, highlighted using various colors, $\bm{\Omega}_{\star}$, $\star \in \lbrace {[\text{Max},i]},{\text{Min}},{[\text{BG},j]} \rbrace$, $i \in \lbrace 1,2,3 \rbrace$, $j \in \lbrace 1,2,3,4 \rbrace$, are explained in more detail in the main text.\label{FIG:MODEL_SETUP}}
	}
\end{figure}

Designing a PTI can be viewed as a problem of determining the material configuration in two PC phases, which, due to their different Bloch-function symmetries, form a topological edge state at their interface. The starting point of our model is a geometric domain split between two phases in a checkerboard topology as shown in Figure \ref{FIG:MODEL_SETUP}\textbf{A}. In a PTI, light impinging from P1 can only scatter to P2 and P4 but not to P3 and thus our hypothesis is that by maximizing the transmission to P2 and P4 while minimizing the transmission to P3, the TO will be forced to generate a PTI. As we will show, this indeed turned out to be the case. We \textcolor{white}{would} like to note that we have also investigated other design domain topologies and find that the inclusion of the sharp bend between the two phases is essential to generate a PTI. Here we consider two PC phases constructed from hexagonal unit cells exhibiting C$_{6\text{v}}$-symmetry \cite{BOOK_SYMMETRY_POWELL}, leaving a twelfth of each hexagonal cell designable for each phase as shown in Figure \ref{FIG:MODEL_SETUP}\textbf{B}. The PC phases are constrained to consist of two materials, silicon and air. The numerical model used in the design procedure detailed in the following considers a spatial domain, $\bm{\Omega} \in \mathbb{R}^2$, cf.\ Figure \ref{FIG:MODEL_SETUP}\textbf{C}. For the design procedure, the two phases are distributed in $\bm{\Omega}$ as shown using black and orange outlines. In this configuration, the PC phases exhibit $180^{\text{o}}$ rotational symmetry around the centre of the hexagonal material slab. 

\begin{figure} 
	\centering
	{
		\includegraphics[width=0.49\textwidth]{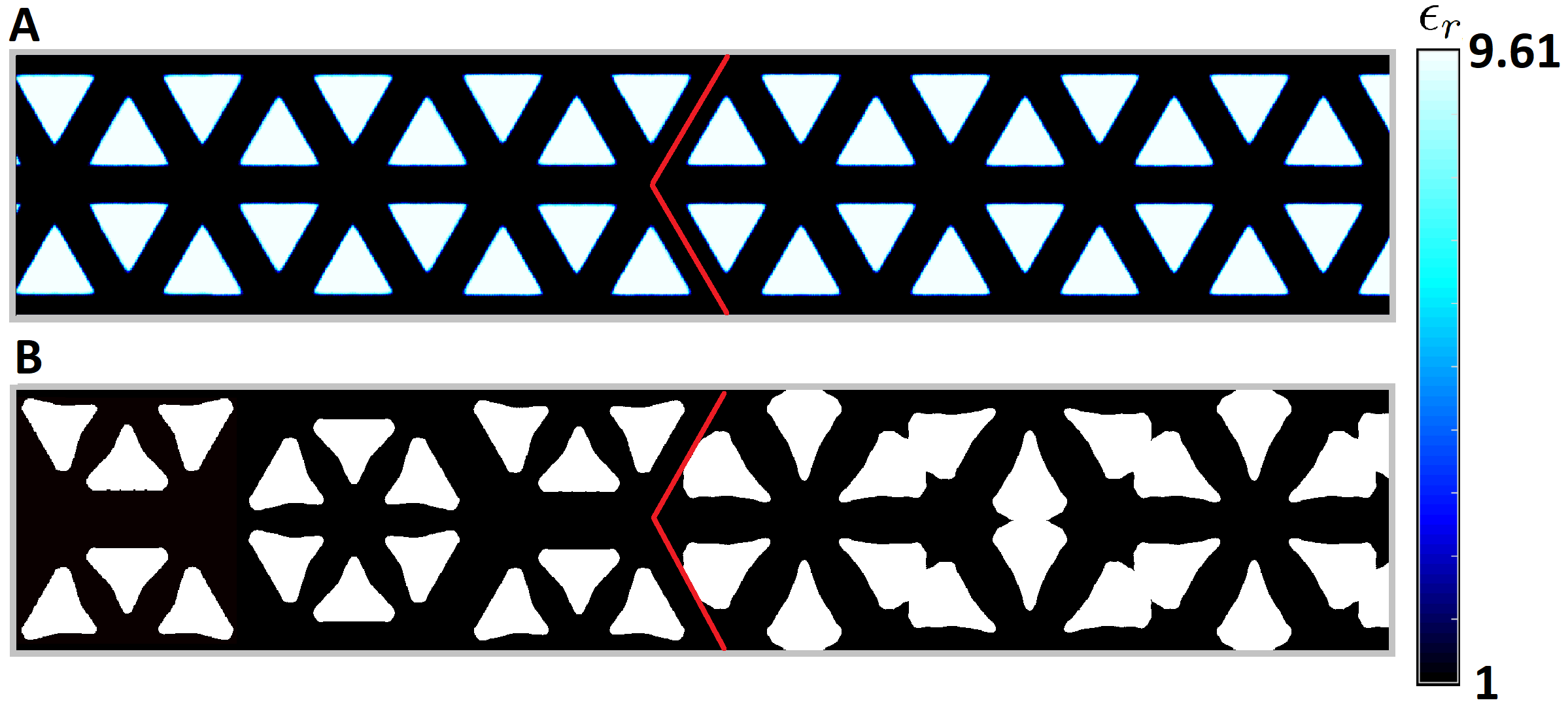} 
		\caption{\textbf{Initial PC and Final PTI Geometry.} \textbf{(A)} The initial material configuration consists of a single PC with triangular air holes (white) in Si (black) for which the two phases are identical. \textbf{(B)} The final PTI material configuration. Both the shape and topology of the second phase has changed from the initial guess while only the shape has changed for the first phase. The interface between the two phases is highlighted in red. \label{FIG:CRYSTAL_STRUCTURES}}
	}
\end{figure} 

We model the light propagation in the frequency domain, 
\begin{equation} \label{EQN:MAXWELL}
\nabla \times \left(\nabla \times \textbf{E}(\textbf{r}) \right) - k_0^2 \epsilon_r(\textbf{r}) \textbf{E}(\textbf{r}) = \textbf{S}(\textbf{r}), \ \ \ \textbf{r} \in \bm{\Omega},
\end{equation}
where $\textbf{E}$ denotes the electric field, $k_0 = \frac{2\pi\nu}{c}$ the free-space wave number with $\nu$ being the frequency and $c$ the speed of light in vacuum, $\epsilon_\text{r}$ is the material-dependent relative permittivity and $\textbf{S}$ denotes the excitation source. The model domain $\bm{\Omega}$ is truncated using a perfectly matched layer (PML)  \cite{Berenger_1994}. In the design process, artificial attenuation is introduced in the subdomain surrounding the hexagonal slab to prohibit the $\textbf{E}$-field from propagating around the outside of the slab from P1 to P2-P4. The attenuation is introduced in Eq.\ \eqref{EQN:MAXWELL} through a complex value of the relative permittivity, $\epsilon_\text{r} = 1 + 10\text{i}$, in the region surrounding the hexagonal slab.

The problem of designing the geometries for the two PC phases is recast as a continuous optimization problem and solved using topology optimization as detailed in \cite{CHRISTIANSEN_2019} and references therein. In short, the transmission from P1 to P2 and P4 is sought maximized for a number of target frequencies while simultaneously minimizing the transmission from P1 to P3 and ensuring that both PC phases have bulk bandgaps at the targeted frequencies. The power flow to P2-P4 as well as inside the bulk of the PC phases for a given excitation $\textbf{S}(\textbf{r})$ and material configuration, $\epsilon_r(\textbf{r})$, is estimated by evaluating the spatially normalized integral of the Poynting vector over $\bm\Omega_{\star}$ defined in Eq. \eqref{EQN:INTENSITY_INTEGRAL}. The target function, $\Phi_{\text{Total}}$, as defined in Eq.\ \eqref{EQN:OPTIMIZATION_PROBLEM} is maximized by (re)distributing silicon and air in the periodic unit cells of each crystal phase, $\bm\Omega_{\text{D},i}, \ i \in \lbrace 1,2 \rbrace$ (see Figure\ \ref{FIG:MODEL_SETUP}\textbf{B}), while ensuring that constraints enforcing bandgaps in the bulk of the PC phases, given in Eq.\ \eqref{EQN:OPTIMIZATION_PROBLEM_CONSTRAINTS}, are satisfied,
\begin{eqnarray}
\max_{\xi(\textbf{r}) \in [0,1]} \ \ \ &\underset{\nu}{\min} \ \ \Phi_{\text{Total}}(\xi)& = \sum_{i=1}^{2} \Phi_{\text{Max},i}(\xi) - \Phi_{\text{Min}}(\xi), \label{EQN:OPTIMIZATION_PROBLEM} \\  
\text{s.t.} \ \ \ &\Phi_{\text{BG,j}}(\xi)& \leq \gamma_{j}, \ \ \ j \in \lbrace 1,2,3,4 \rbrace  \label{EQN:OPTIMIZATION_PROBLEM_CONSTRAINTS}
\end{eqnarray}
with,
\begin{eqnarray} \label{EQN:INTENSITY_INTEGRAL}
\Phi_{\star}(\nu,\xi) &=& \left. \tau_{\star} \int_{\bm{\Omega}_{\Phi_{\star}}} \vert \text{P}(\textbf{E}(\nu,\xi)) \vert \text{d}\textbf{r} \middle/ \int_{\bm{\Omega}_{\Phi_{\star}}} \text{d}\textbf{r} \right. , \\ 
\star \in \lbrace {[\text{Max},i]},&{\text{Min}}&,{[\text{BG},j]} \rbrace, i \in \lbrace 1,2,3 \rbrace, j \in \lbrace 1,2,3,4 \rbrace. \nonumber
\end{eqnarray}
Here $\textbf{P}$ denotes the Poynting vector and $\tau_{\star}$ a set of scaling constants. The subdomains $\bm{\Omega}_{\Phi_{\star}}$ are all shown in Figure\ \ref{FIG:MODEL_SETUP}\textbf{C}. 

The target function consists of three contributions: $\Phi_{\text{Max,1}}$ measuring the power flow to P2 and P4 (the region $\bm\Omega_{\text{Max},1}$), $\Phi_{\text{Max,2}}$ measuring the power flow at the centre of the hexagonal slab (the region $\bm\Omega_{\text{Max},2}$), and $\Phi_{\text{Min}}$, which measures the power flow to P3 (the region $\bm\Omega_{\text{Min}}$). The four constraint functions: $\Phi_{\text{BG,j}}(\xi), \ \ \ j \in \lbrace 1,2,3,4 \rbrace$ are computed by evaluating the power flow in the regions $\bm\Omega_{\text{BG},j}$. The choice of the target function in Eq.\ \eqref{EQN:OPTIMIZATION_PROBLEM} and the configuration of $\bm{\Omega}_{\text{Max},i}$ and $\bm{\Omega}_{\text{Min}}$ assures that by solving the optimization problem the vast majority of the power coupled into the hexagonal slab at P1 will flow to P2 and/or P4 with only minimal power flowing to P3. The inclusion of $\Phi_{\text{Max,2}}$ in the target function ensures that light propagates to the center of the domain in the initial steps of the optimization but has negligible influence on the final result. The choice of Eq. \eqref{EQN:OPTIMIZATION_PROBLEM_CONSTRAINTS} and $\bm{\Omega}_{\text{BG},j}$ assures that only a negligible amount of power flows into the bulk of the PC phase, effectively ensuring bulk bandgaps in both PC phases at the targeted frequencies. 

\section{3. Results}

We model a silicon slab perforated with air holes, which is implemented in a 2D using a reduced effective permittivity of $\epsilon_{r_{\text{Si}}} = 9.61$ for silicon while $\epsilon_{r_{\text{air}}} = 1$ is assumed for the air. A lattice constant of $a = 735$ nm is chosen for the hexagonal unit cells, cf.\ Figure \ref{FIG:MODEL_SETUP}\textbf{A}. The excitation, $\textbf{S}(x)$, is chosen to be a transverse electric (TE) polarized point source. The design problem defined by Eqs.\ \eqref{EQN:MAXWELL}-\eqref{EQN:OPTIMIZATION_PROBLEM_CONSTRAINTS} is solved for the two frequencies $\nu_1 = 184$ THz and $\nu_2 = 187$ THz (approximately corresponding to the wavelengths $\lambda_1 = 1630$ nm, $\lambda_2 = 1604$ nm) using a min/max approach \cite{BOOK_TOPOPT_BENDSOE}. The initial material distribution is chosen to be a photonic crystal structure with equilateral triangular holes and a bandgap in the considered frequency range, inspired by the recent work of Barik et al. \cite{BARIK_2016}, see Figure \ref{FIG:CRYSTAL_STRUCTURES}\textbf{A}. Note that the starting material configuration includes a smooth transition in the dielectric function around the edges, which is a \textcolor{white}{result} of the filtering process used in the TO algorithm, see e.g. \cite{CHRISTIANSEN_SIGMUND_2016}.

The material distributions for both PC phases resulting from solving the design problem are shown in Figure \ref{FIG:CRYSTAL_STRUCTURES}\textbf{B}. The interface between the phases is highlighted in red, and black/white corresponds to silicon/air, respectively. From the figure it is clearly seen that the shape of the air holes in the left PC phase has changed as a result of the design process, while for the right PC phase both a shape change and a change of the topology of the air holes has occurred. 

\begin{figure}[h] 
	\centering
	{
		\includegraphics[width=0.495\textwidth]{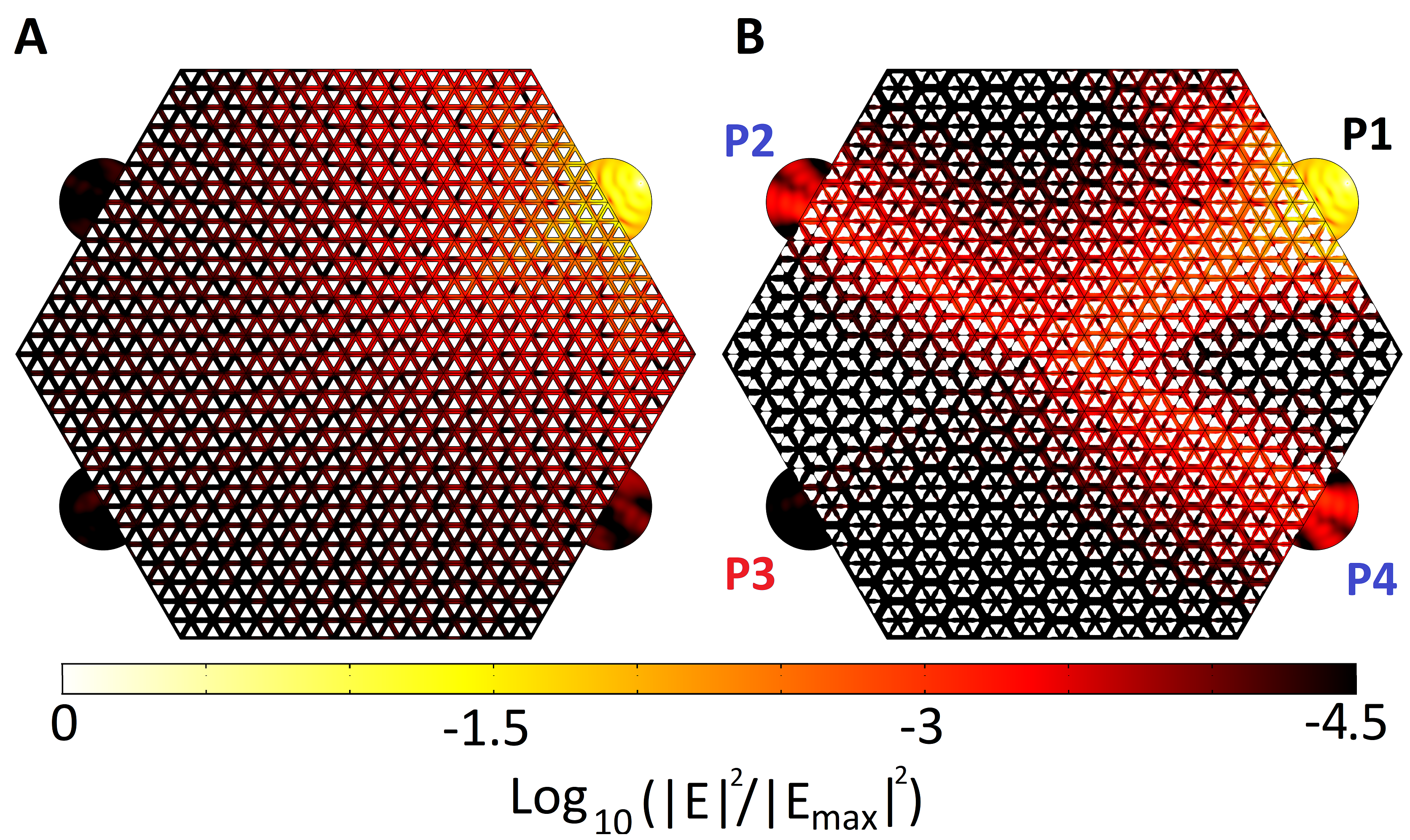}
		\caption{\textbf{Energy distribution in design domain.} The max-normalized energy distribution in the design setup for the initial crystal (\textbf{A}) and final PTI crystal configuration (\textbf{B}) shows a transmission improvement to ports P2 and P4 by orders of magnitude. The slab is excited by a TE-polarized point source at P1 oscillating at 184 THz. \label{FIG:HEXAGONAL_CONFIGURATION}}
	}
\end{figure}

To demonstrate the topological behaviour of the designed PTI, Figure\ \ref{FIG:HEXAGONAL_CONFIGURATION} shows the max-normalized electric field intensity on a logarithmic scale, for the chosen excitation at $\nu = 184$ THz ($\lambda = 1630$ nm), for the initial and final material distributions. From Figure\ \ref{FIG:HEXAGONAL_CONFIGURATION}\textbf{A} it is clearly observed that the initial PC configuration exhibits a bandgap at 184 THz with the energy dropping several orders of magnitude as the field extends into the crystal. In contrast, for the topological edge-state generated by our TO design shown in Figure\ \ref{FIG:HEXAGONAL_CONFIGURATION}\textbf{B}, the vast majority of the electric field entering at P1 propagates to either P2 or P4 while very little energy propagates to P3 ($\approx 1\%$), indicating the successful design of a back-scattering-suppressing structure. 

\begin{figure*}[!]
	\centering
	{
		\includegraphics[width=0.97\textwidth]{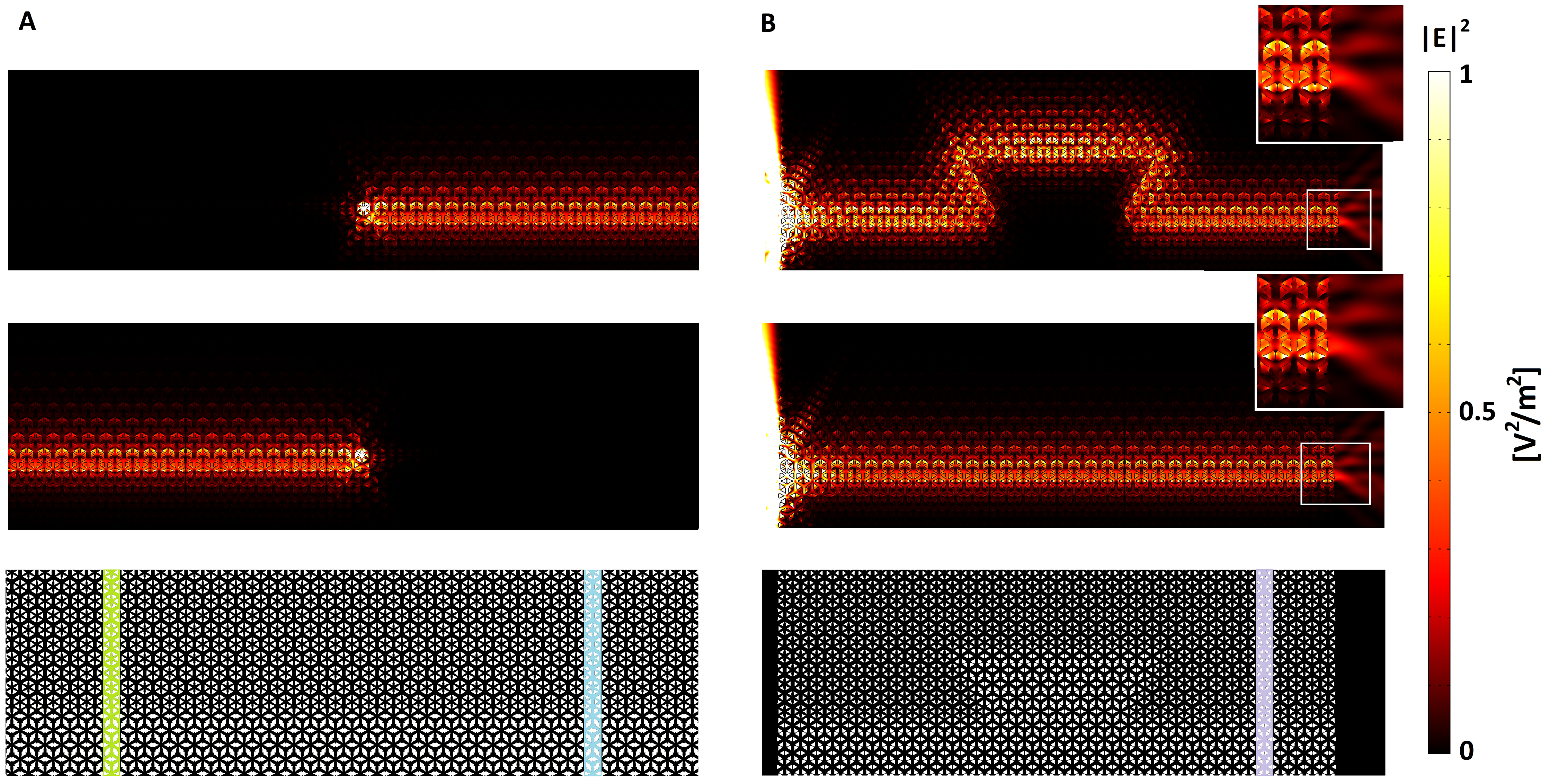} 
		\caption{\textbf{Spin-dependent directional emission and robust propagation past $120^{\text{o}}$ bends.} \textbf{(A)} The PTI is excited using a chiral source emitting at 184 THz placed near the interface and rotating clockwise (top) and counter-clockwise (middle) for the material distribution shown in the bottom panel. The emission direction depends strongly on the chirality of the emitter, which clearly demonstrates spin-dependent directional emission for emitters placed at the PTI interface. \textbf{(B)} Field propagation at 184 THz through four $120^{\text{o}}$ bends (top) for the material distribution shown in the bottom panel compared to propagation along a straight channel (middle). This demonstrates the back-scattering-robust propagation along the PTI interface. The color map is truncated at $\vert \textbf{E} \vert^2 = 1 \ \ \text{V}^2/\text{m}^2$ as the field-strength at the source diverges. \label{FIG:BACK_SCATTERING_ROBUSTNESS}}
	}
\end{figure*}

Chiral quantum optical effects are pronounced in PTI edge states due to the spin-momentum locking of the quantum-spin-Hall states. This effect is very prominent in our geometry, as shown in Figure\ \ref{FIG:BACK_SCATTERING_ROBUSTNESS}\textbf{A}, which shows the emission intensity from emitters with opposite chirality (spin), placed at $(x_s,y_s) = (0 \text{nm},70 \text{nm})$ relative to the centre of the unit cell of Phase 1 closest to the interface. By extracting the emitted intensity going left or right for a given emitter chirality, we extract directional $\beta$-factors for left- and right-hand circularly polarized emitters as
\begin{eqnarray}
\beta_k = \frac{ \int \vert \textbf{E} \vert^2 \text{d}\bm{\Omega}_{k} }{ \int \vert \textbf{E} \vert^2 \text{d}\left(\bm{\Omega}_{\text{L}} \bigcup \bm{\Omega}_{\text{R}}\right) }, \ \ k \in \lbrace \text{L,R} \rbrace,
\end{eqnarray}
with $\bm{\Omega}_{L}$ and $\bm{\Omega}_{R}$ highlighted in the bottom panel of Figure\ \ref{FIG:BACK_SCATTERING_ROBUSTNESS}\textbf{A} using green and blue respectively. We exctract $\beta_\text{L} = \beta_\text{R} > 0.995$, corresponding to a perfectly deterministic directional light-matter interface \cite{PAP_Lodahl2017} within the numerical error of our calculations. 

As a demonstration of the back-scattering-suppressed field propagation past sharp bends offered by the PTI, two additional slab models are constructed. The first model consists of a slab with four 120$^{\text{o}}$ bends along the interface between the two PC phases, the geometry is shown in the bottom panel of Figure\ \ref{FIG:BACK_SCATTERING_ROBUSTNESS}\textbf{B}. The second model consists of a straight line interface between the two PC phases identical to the one shown in the bottom panel of Figure\ \ref{FIG:BACK_SCATTERING_ROBUSTNESS}\textbf{A}. For both models, a point source emitting a TE polarized field at 184 THz is placed outside the slab on the left hand side and aligned with the PC phase interface position. The resulting electric-field intensities in the slabs are shown in the top and middle panels of Figure\ \ref{FIG:BACK_SCATTERING_ROBUSTNESS}\textbf{B}. The top panel shows the slab with bends while the middle shows the slab with the straight interface. An inset showing a magnified image of the output is included at the top right of both panels, showing qualitatively identical intensity. To obtain a quantitative measure of the loss from traversing the four bends, the integral of $\vert \textbf{E} \vert^2$ in the regions highlighted in blue in the bottom panel of Figure\ \ref{FIG:BACK_SCATTERING_ROBUSTNESS}\textbf{A} and purple in the bottom panel of Figure\ \ref{FIG:BACK_SCATTERING_ROBUSTNESS}\textbf{B}, is computed. Hereby, we find that the loss is less than 0.04 dB per bend. 

\begin{figure*}[!]
	\centering
	{
		\includegraphics[width=0.97\textwidth]{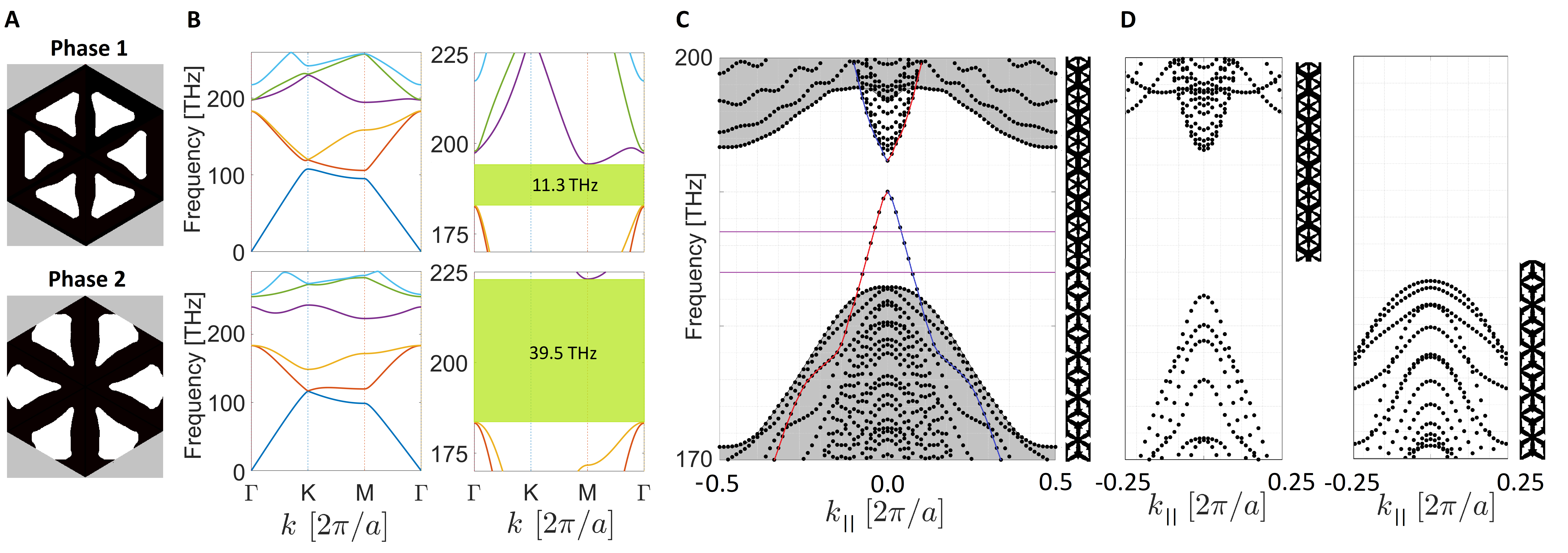} 
		\caption{\textbf{Geometries and band structures of PTIs.} \textbf{(A)} The two PC phases constituting the PTI. \textbf{(B)} Band structure diagrams for each individual PC phase showing the first six bands. \textbf{(C)} Band structures for the PTI supercell on the right \textcolor{white}{with the pseudo-spin edge modes highlighted using red and blue and the frequencies targeted in the topology optimization process overlaid as horizontal magenta lines}. \textbf{(D)} Band structures for the supercells, on the right, constructed from each PC crystal phase showing the appearance of the artificial standing wave Bloch modes in each phase seen in the PTI supercell band structure in \textbf{(C)}. \label{FIG:BAND_STRUCTURES}}
	}
\end{figure*} 

Finally, we have calculated the band structures of both PC phases as well as the dispersion relation of the edge state. The result is presented in Figure\ \ref{FIG:BAND_STRUCTURES} where Figure\ \ref{FIG:BAND_STRUCTURES}\textbf{A} shows a unit cell from each of the two PC phases while Figure\ \ref{FIG:BAND_STRUCTURES}\textbf{B} shows the associated band structures for the lowest six bands. The left column shows the full band structures and the right column shows a zoom around the target frequencies with an overlay highlighting the bulk bandgaps. From the band structure for phase 1 it is seen that bands 2 and 3 as well as bands 4 and 5 experience a degeneracy at the $\Gamma$-point. For phase 2, a degeneracy is only observed at the $\Gamma$-point for bands 2 and 3 whereas band 4 has a bandgap on both sides. The degeneracies at the $\Gamma$-point \textcolor{white}{form} the foundation for the formation of the PTI edge state as the two PC phases are brought into contact. Figure\ \ref{FIG:BAND_STRUCTURES}\textbf{C} shows the band diagram for the supercell, obtained by imposing periodic boundary conditions on the left/right of the supercell and Neumann boundary conditions on the top/bottom. \textcolor{white}{The frequencies targeted in the topology optimization process are included as horizontal magenta lines.} The introduction of the Neumann boundary conditions leads to a number of artificial standing wave Bloch-modes becoming solutions to the eigenvalue problem solved to obtain the band structure. To verify that these modes are indeed artificial Bloch-modes, we have calculated band diagrams for each of the two PTI phases as shown in Figure\ \ref{FIG:BAND_STRUCTURES}\textbf{D} and observe that these modes clearly are numerical artefacts stemming from the resonances in each of the two phases. By filtering these artificial modes away, one arrives at a pair of counter-propagating edge modes inside the bulk-bandgap region highlighted in Figure\ \ref{FIG:BAND_STRUCTURES}\textbf{C} using blue and red according to the pseudo-spin associated with each mode. The horizontal purple lines have been added in the diagram to denote the frequencies at which the PTI was designed. The band structure in Figure\ \ref{FIG:BAND_STRUCTURES}\textbf{C} clearly shows that interface edge-states exist in the bulk bandgap for the two PC phases. A small bandgap is found in the edge-states, which may be attributed to the interface itself breaking the C$_\text{6v}$ symmetry \cite{Parappurath2018}. The PTI phases have a central frequency of 188.6 THz (203.3 THz) and bandgap of 11.3 THz (39.5 THz) for Phase 1 (Phase 2), which corresponds to relative bandgap widths of 0.06 and 0.19, respectively. This is 50 \% or more larger than the relative bandgaps found by Barik et al. \cite{BARIK_2016}, which directly shows the power of employing TO to design PTIs.

\section{4. Conclusion}
We have used inverse design by TO to generate a PTI structure without imposing any geometrical constraints beyond the C$_\text{6v}$ symmetry. It is interesting to note that although our algorithm is set up to maximize the transmission of light through a crystal with two different phases including a sharp bend, a number of interesting features and phenomena emerge as indirect benefits, such as larger band gaps than previous designs, extremely low bending losses, as well as (within the numerical error) unity directionality of embedded emitters. Neither of these were design constraints but simply emerge as a result of the TO, which is a direct testimony to the power of TO when applied to complex inverse problems. Notably, it would be rather trivial to extend our formulation to, e.g., also target a widening of the bandwidth by introducing more target frequencies which are spread further apart but perhaps even more importantly, our method may be employed to generate entirely new PTI designs based on different crystal symmetries and constraints, which is an exciting prospect for further work.

\begin{acknowledgments}
The authors acknowledge support from NATEC (NAnophotonics for Terabit Communications) Centre (Grant No. 8692) and the Villum Foundation Young Investigator Programme.
\end{acknowledgments}

\bibliographystyle{apsrev4-1} 
\bibliography{References}   

\end{document}